\documentclass[a4paper,11pt]{article}
\usepackage{pos}
\usepackage{amsmath}
\usepackage{xspace}
\usepackage{tikz}
\usetikzlibrary{positioning}

\newcommand{\rabbit}{\textsc{Rabbit}\xspace}
\newcommand{\combine}{\textsc{Combine}\xspace}
\newcommand{\combinetf}{\textsc{CombineTF}\xspace}
\newcommand{\pyhf}{\texttt{pyhf}\xspace}
\newcommand{\tensorflow}{TensorFlow\xspace}
\newcommand{\jax}{JAX\xspace}
\newcommand{\scipy}{SciPy\xspace}
\newcommand{\cpp}{C\texttt{++}\xspace}
\newcommand{\nbins}{N^{\mathrm{bins}}}
\newcommand{\nproc}{N^{\mathrm{proc}}}
\newcommand{\nsyst}{N^{\mathrm{syst}}}

\title{Efficient binned profile likelihood minimization for precision measurements with RABBIT}
\ShortTitle{Efficient binned profile likelihood minimization with RABBIT}

\author*[a]{David Walter}
\author[b]{Josh Bendavid}
\author[c]{Kenneth Long}

\affiliation[a]{Massachusetts Institute of Technology,\\
  77 Massachusetts Avenue, Cambridge, MA, USA}

\affiliation[b]{European Organization for Nuclear Research (CERN),\\
  Esplanade des Particules 1, Geneva, Switzerland}

\affiliation[c]{Universit\'e Claude Bernard Lyon 1, CNRS/IN2P3, IP2I Lyon,\\
  Villeurbanne, France}

\emailAdd{david.walter@cern.ch}
\emailAdd{josh.bendavid@cern.ch}
\emailAdd{kenneth.long@cern.ch}

\abstract{Precision measurements at the LHC increasingly rely on binned profile maximum likelihood fits with thousands of bins and nuisance parameters, and the High-Luminosity LHC will push these numbers further. Fast and robust minimization of such likelihoods is crucial for timely analysis development and accurate inference. We present \rabbit (Rapid Automatic Bin-Based Inference Tool), a Python framework that exploits differentiable programming in TensorFlow~2 to perform this task on CPUs and GPUs. Automatic differentiation provides exact gradients and Hessian-vector products for a trust-region minimizer operating in Krylov subspaces, and just-in-time compilation yields near-\cpp execution speed. \rabbit implements flexible statistical models with analytic treatments where possible, supports symmetrization options that establish Gaussian approximations and a linearized likelihood formulation with deterministic solutions, and focuses on measuring physical observables through differentiable transformations of the model, including unfolded differential cross sections. Benchmarks on synthetic models demonstrate excellent scaling with the number of bins and parameters, outperforming established tools in challenging regimes where these fail to converge within reasonable time.}

\FullConference{23rd International Workshop on Advanced Computing and Analysis Techniques in Physics Research (ACAT2025)\\
8--12 September 2025\\
Hamburg, Germany\\}

\begin{document}
\maketitle

\section{Introduction}
\label{sec:intro}

In binned profile maximum likelihood analyses, parameters of interest (POIs) are extracted by adjusting templates of the expectation, which are functions of the POIs and of numerous nuisance parameters encoding systematic effects, to describe the observed histograms. Maximizing sensitivity requires fine binning across multiple observables, while controlling systematic effects demands a detailed decomposition of uncertainty sources: the CMS measurement of the W boson mass based on a fraction of the LHC Run~2 data already uses over 3\,000 bins and 4\,000 nuisance parameters~\cite{mw}, numbers that could grow by orders of magnitude with the full Run~2, Run~3, and especially High-Luminosity LHC datasets.

\rabbit (Rapid Automatic Bin-Based Inference Tool)~\cite{rabbit} is a flexible and efficient framework for this task built on top of TensorFlow~2~\cite{tensorflow}. Its performance derives from multi-threaded CPU and GPU acceleration, automatic differentiation for exact gradient and Hessian calculations, just-in-time (JIT) compilation, an improved minimization strategy, and strict vectorization. \rabbit further reduces the effective complexity of likelihoods through analytic solutions and a linearized formulation, and it disentangles the likelihood parameterization from the quantities of interest: physical observables are obtained through differentiable transformations of the model, handled automatically. \rabbit thereby complements existing frameworks commonly used in statistical analysis at the LHC such as \combine~\cite{combine} and \pyhf~\cite{pyhf} with better scaling behavior for complex measurements.

\section{Statistical model and numerical optimization}
\label{sec:model}

The statistical model combines an observation model, constrained parameters for systematic uncertainties, bin-by-bin parameters for the statistical uncertainty of the prediction, and an optional regularization term. For observed counts $n^{\mathrm{obs}}_i$ in $\nbins$ bins, the default negative log-likelihood reads, omitting constant terms,
\begin{equation}
\label{eq:nll}
    L = \sum_{i} \left[ n_i - n^{\mathrm{obs}}_i \ln(n_i) \right]
      + \frac{1}{2} \sum_{k} \left(\theta_k - \theta^0_k\right)^2
      + \sum_{i,j} k^{\mathrm{stat}}_{i,j} \left[ \beta_{i,j} - \beta^0_{i,j} \ln(\beta_{i,j}) \right]
      + L_\mathrm{reg}\,,
\end{equation}
where $i$, $j$, and $k$ run over the $\nbins$ bins, $\nproc$ processes, and $\nsyst$ systematic uncertainties, respectively. The first term is the Poisson observation model with expected yields $n_i = \sum_{j} \mu_j\, \beta_{i,j}\, n^{0}_{i,j} \prod_{k} \kappa_{\pm,i,j,k}^{\theta_k}$, where the signal-strength modifiers $\mu_j$ are unconstrained for signals and unity for backgrounds; $\chi^2$ observation models with diagonal or non-diagonal covariance are also available. The second term imposes Gaussian constraints on the nuisance parameters $\theta_k$, which scale the systematic effects multiplicatively through the $\kappa$ tensor, yielding log-normally distributed variations; an additive treatment, $n_i \rightarrow n_i + \sum_k v_{\pm,i,k}\,\theta_k$, yielding normally distributed variations and a (partially) linearized likelihood, is also supported. Asymmetric variations are interpolated with smooth piecewise polynomials as in Ref.~\cite{combine}, or symmetrized with one of several options (average, conservative, linear, quadratic) that simplify the optimization and improve the validity of the Gaussian approximation.

The third term accounts for the statistical uncertainty of the prediction with the Barlow--Beeston method~\cite{barlow_beeston}, shown in its general gamma-distributed form with one parameter $\beta_{i,j}$ per bin and process (``full''), where $k^{\mathrm{stat}}_{i,j} = (n^{0}_{i,j})^2/(\Delta n^{0}_{i,j})^2$ is the effective sample size and $\beta^0_{i,j}$ is nominally unity; a single effective parameter per bin (``lite'')~\cite{barlow_beeston_lite}, as well as normally distributed constraints, are also supported. Wherever possible, these parameters are profiled analytically and do not enter the numerical optimization explicitly. Finally, $L_\mathrm{reg}$ is an optional regularization term that users can configure, e.g.\ for regularized unfolding.

All gradients, Hessians, and Hessian-vector products (HVPs) are computed exactly with automatic differentiation, and the entire pipeline is JIT-compiled, giving near-\cpp speed, kernel fusion that avoids materializing intermediate results in memory, and automatic utilization of available CPU cores or GPUs. The minimization interfaces \scipy~\cite{scipy} optimizers, primarily a trust-region algorithm operating in Krylov subspaces~\cite{trust_krylov} that solves the trust-region subproblem from HVPs rather than full Hessian matrices, reducing cost and memory while handling ill-conditioned problems. The full Hessian is computed once after convergence to estimate the parameter covariance and the expected distance to the minimum, used to validate the fit.

When all systematic uncertainties are symmetrized and act additively, and a $\chi^2$ observation model is used, the likelihood becomes purely quadratic: the maximum likelihood estimate is obtained deterministically in a single step, $\hat{\mathbf{x}} = \mathbf{x} + \mathbf{H}^{-1}\mathbf{J}$, optionally with a Hessian-free conjugate-gradient solve, or fully analytically by absorbing systematic effects into the covariance matrix when only unconstrained parameters are of interest.

\section{Workflow and user interface}
\label{sec:workflow}

\begin{figure}[t]
\centering
\begin{tikzpicture}[
  box/.style={draw, rounded corners, align=center, font=\small, inner sep=4pt, minimum height=1.1cm},
  lab/.style={align=center, font=\footnotesize\itshape}
]
\node[box] (in) {Input histograms (UHI)\\ data, pseudodata, predictions,\\ systematic variations};
\node[box, right=1.9cm of in] (tensor) {\rabbit tensor\\ (HDF5)};
\node[box, right=1.9cm of tensor] (res) {Results (HDF5)\\ parameters, impacts, scans,\\ post-fit distributions};
\draw[-stealth, thick] (in) -- node[lab, above=2pt] {Python\\ interface} (tensor);
\draw[-stealth, thick] (tensor) -- node[lab, above=2pt] {command-line\\ fitting tools} (res);
\end{tikzpicture}
\caption{The two-stage \rabbit workflow: a Python interface assembles input histograms into the \rabbit tensor, which is stored and then processed by command-line tools for minimization and inference.}
\label{fig:workflow}
\end{figure}
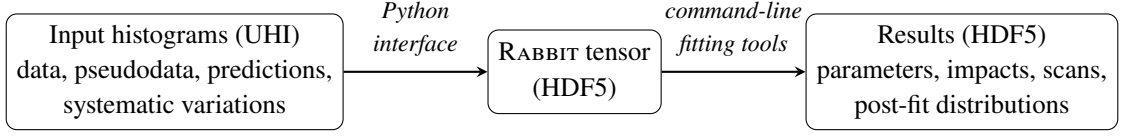

A two-stage workflow, sketched in Figure~\ref{fig:workflow}, separates data preparation from inference. A lightweight Python interface assembles observations, optional pseudodata, and per-process expectations, provided as histograms implementing the Unified Histogram Interface~\cite{uhi} (e.g.\ boost histograms~\cite{boost}), into the \rabbit tensor, serialized to an HDF5 file. Histograms from different event selections or datasets are combined as separate channels, with the original axes preserved in metadata for channel- and axis-aware transformations of the results. Systematic uncertainties are specified by one (symmetric) or two (asymmetric) histograms that express the modification of the expected yields under a variation of the corresponding source, with the symmetrization and the multiplicative or additive treatment chosen per uncertainty; a conversion script accepts inputs in \combine datacard format. A command-line interface then loads the tensor and performs the minimization together with auxiliary computations such as impacts, post-fit distributions, likelihood scans, goodness-of-fit tests, toys, or Asimov fits.

\section{Features and applications}
\label{sec:features}

\paragraph{Mappings and masked channels.}
Mappings are arbitrary differentiable transformations of the expected yields, nuisance parameters, and signal strengths, such as projections, normalized distributions, ratios, and asymmetries, evaluated with full post-fit uncertainty propagation; custom mappings can be provided by the user. Masked channels are auxiliary distributions that are part of the model but excluded from the likelihood, as no observed data is associated with them. They give access to derived quantities together with their full uncertainty. A measured cross section, for example, depends not only on the fitted signal strength but also on the nuisance parameters through $\hat{\sigma} = \hat{\mu}\,\sigma(\hat{\theta})$. By placing the predicted (differential) cross section and its systematic variations in a masked channel, \rabbit propagates all post-fit parameters to evaluate it with the correct uncertainty, which is the basis for unfolding applications.

\paragraph{Uncertainty decomposition and likelihood inference.}
Traditional nuisance-parameter impacts, shifting each parameter by $\pm1$ standard deviation of its post-fit uncertainty, are obtained in the linear approximation directly from the parameter covariance; alternatively, impacts of global observables~\cite{globalImpacts} shift the external constraint values, with the advantage that individual and grouped impacts add in quadrature, which is beneficial for the combination of independent measurements. Beyond the Gaussian approximation, profiled likelihood scans are available in one and two dimensions, and confidence-level contours can be scanned directly by maximizing the radius in the space of the observables of interest at fixed $-2\Delta\ln\mathcal{L}$. Goodness of fit is quantified by a $\chi^2$ test on the residuals of any physics model and by the likelihood difference to the saturated model~\cite{saturated}; asymptotic CLs limits~\cite{cls_junk,cls_read} and a deterministic parameter-blinding procedure are also provided.

\paragraph{Recent developments.}
Since the workshop, development has focused on parameterized models in which process yields are arbitrary differentiable functions of the fit parameters, including normalization, exponential, and Bernstein-polynomial parameterizations of histogram axes, and composite combinations thereof. These enable in-situ data-driven background estimation, e.g.\ ABCD methods with optional Chebyshev-polynomial smoothing to reduce the number of free parameters. External likelihood terms with user-provided gradients and (sparse) Hessians can also be added, e.g.\ for external constraints or combinations.

\section{Validation and benchmarks}
\label{sec:benchmark}

The implementation has been validated against \combine~\cite{combine}: negative log-likelihood values and likelihood scans agree exactly for equivalent model configurations, after accounting for constant offsets.

Performance is evaluated on synthetic models with one signal and four background processes, 200 events per process per bin, and bin-correlated systematic variations; pseudodata are generated with Poisson fluctuations around randomly sampled parameters. All benchmarks minimize a Poisson likelihood with log-normal constraints and gamma-distributed Barlow--Beeston-lite uncertainties, followed by uncertainty estimation in the Gaussian approximation; a fit passes if it converges within 10 minutes. Tests were run on the SubMIT facility at MIT~\cite{submit}, on servers with dual AMD EPYC 9654 (368 threads) and 9965 (768 threads) CPUs with 1.5\,TB of RAM, and on an NVIDIA Tesla V100 GPU (32\,GB).

\begin{figure}[t]
\centering
\includegraphics[width=0.37\textwidth]{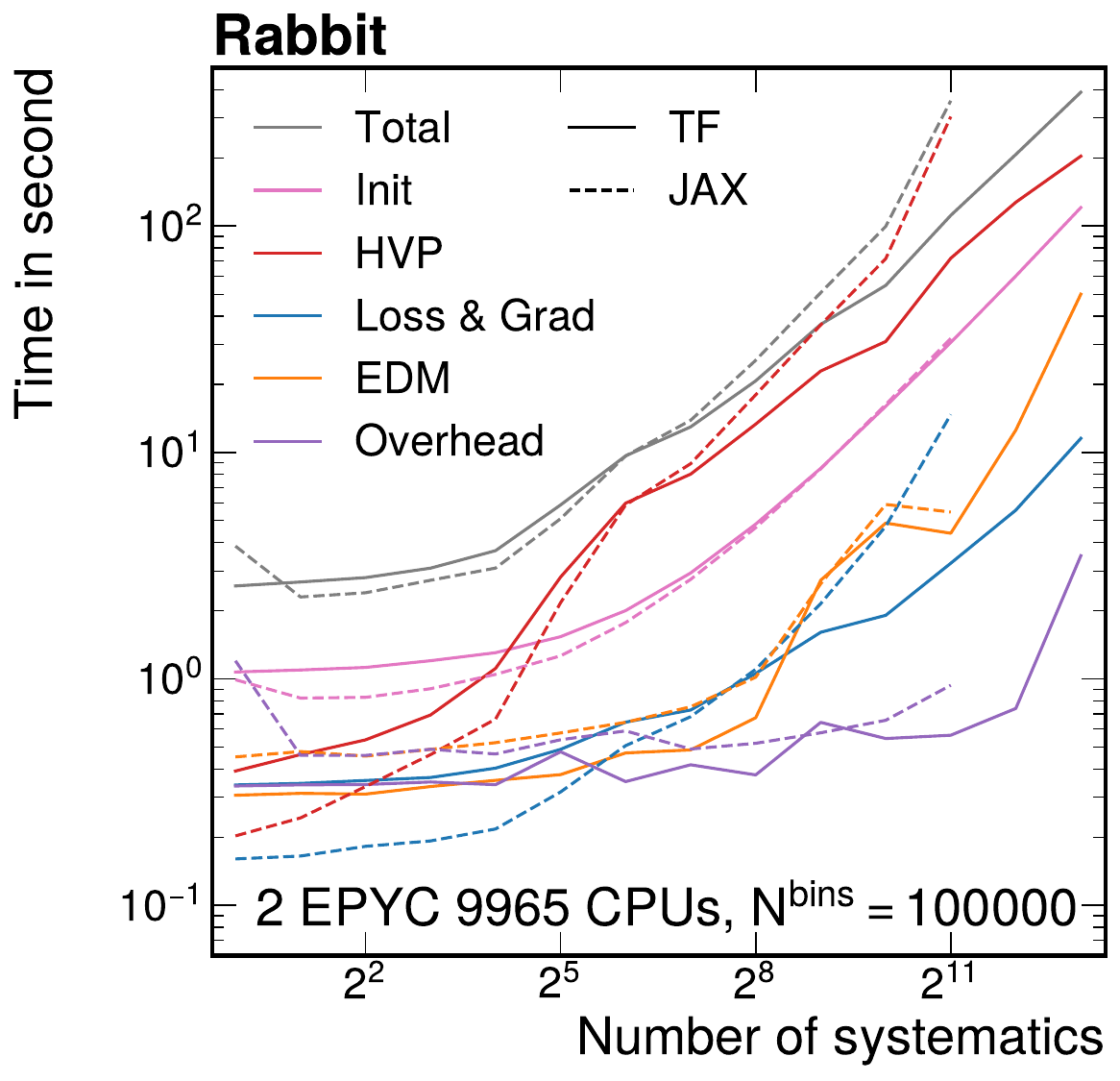}
\includegraphics[width=0.37\textwidth]{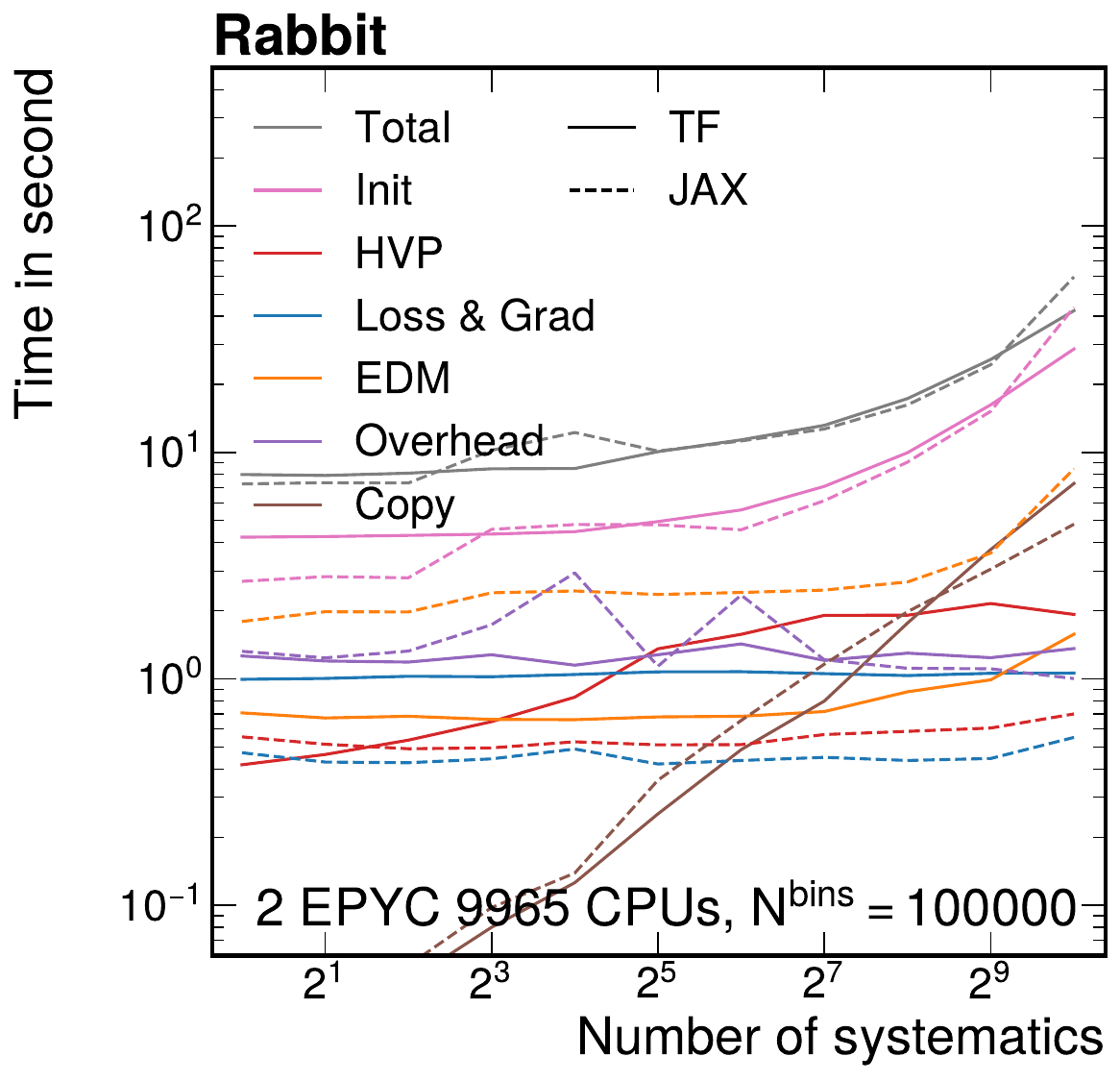}
\caption{Fit time for a model with 100\,000 bins versus the number of systematic uncertainties, broken down into computational steps, for the \tensorflow (solid) and \jax (dashed) backends on a dual AMD EPYC 9965 server (left) and a Tesla V100 GPU (right).}
\label{fig:timing}
\end{figure}

Figure~\ref{fig:timing} shows the runtime breakdown for models with $\nbins = 100\,000$: approximately constant for small models, where initialization and JIT compilation dominate, and growing roughly linearly with complexity. On CPUs the HVP evaluations in the minimizer dominate for complex models, whereas on GPUs automatic differentiation is substantially faster and initialization dominates; the largest model fitted on the GPU had $\nsyst = 2\,048$, limited by GPU memory. An alternative \jax-based implementation shows largely compatible performance; only \tensorflow is retained as the supported backend.

\begin{figure}[t]
\centering
\includegraphics[width=0.37\textwidth]{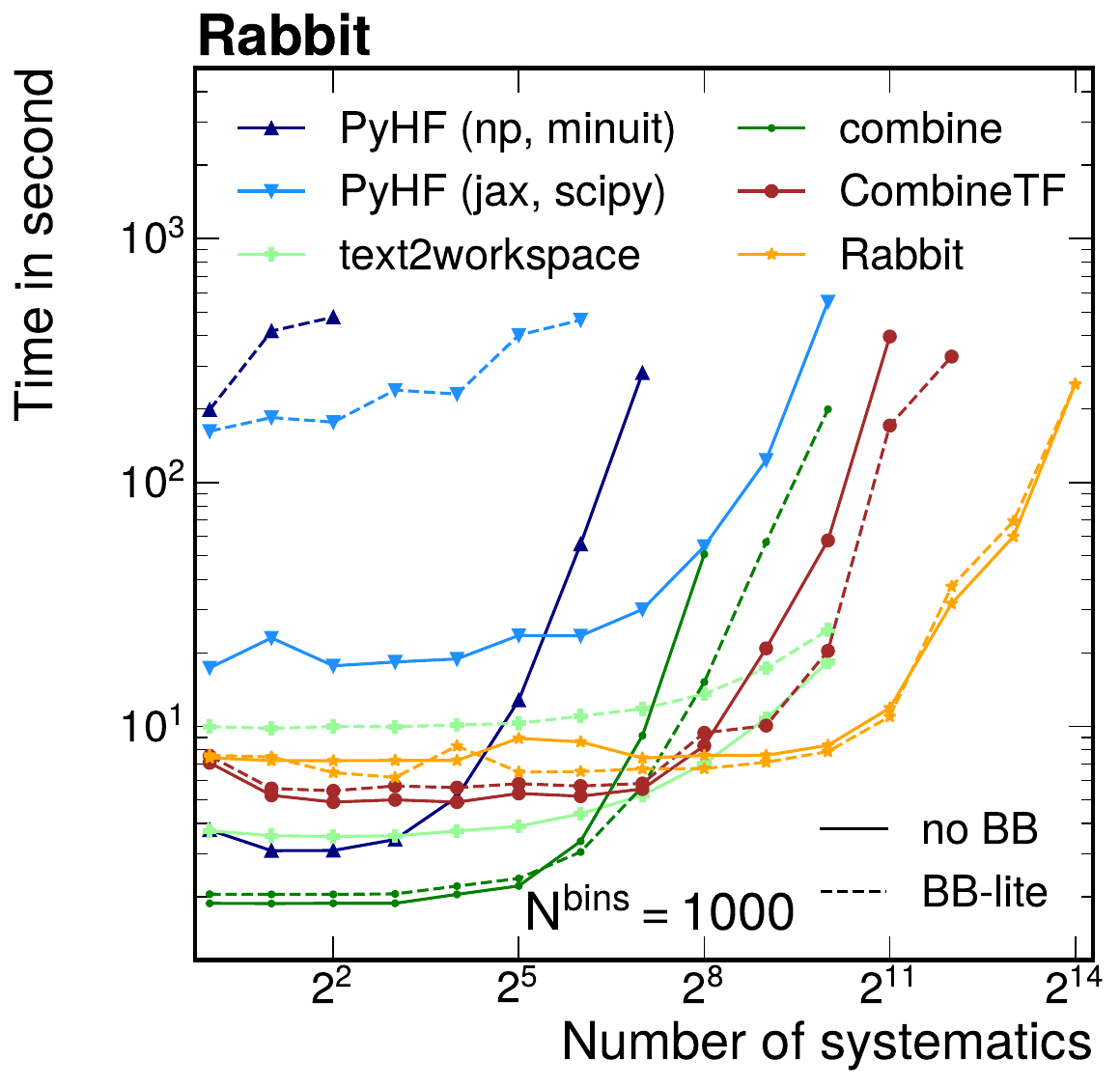}
\includegraphics[width=0.37\textwidth]{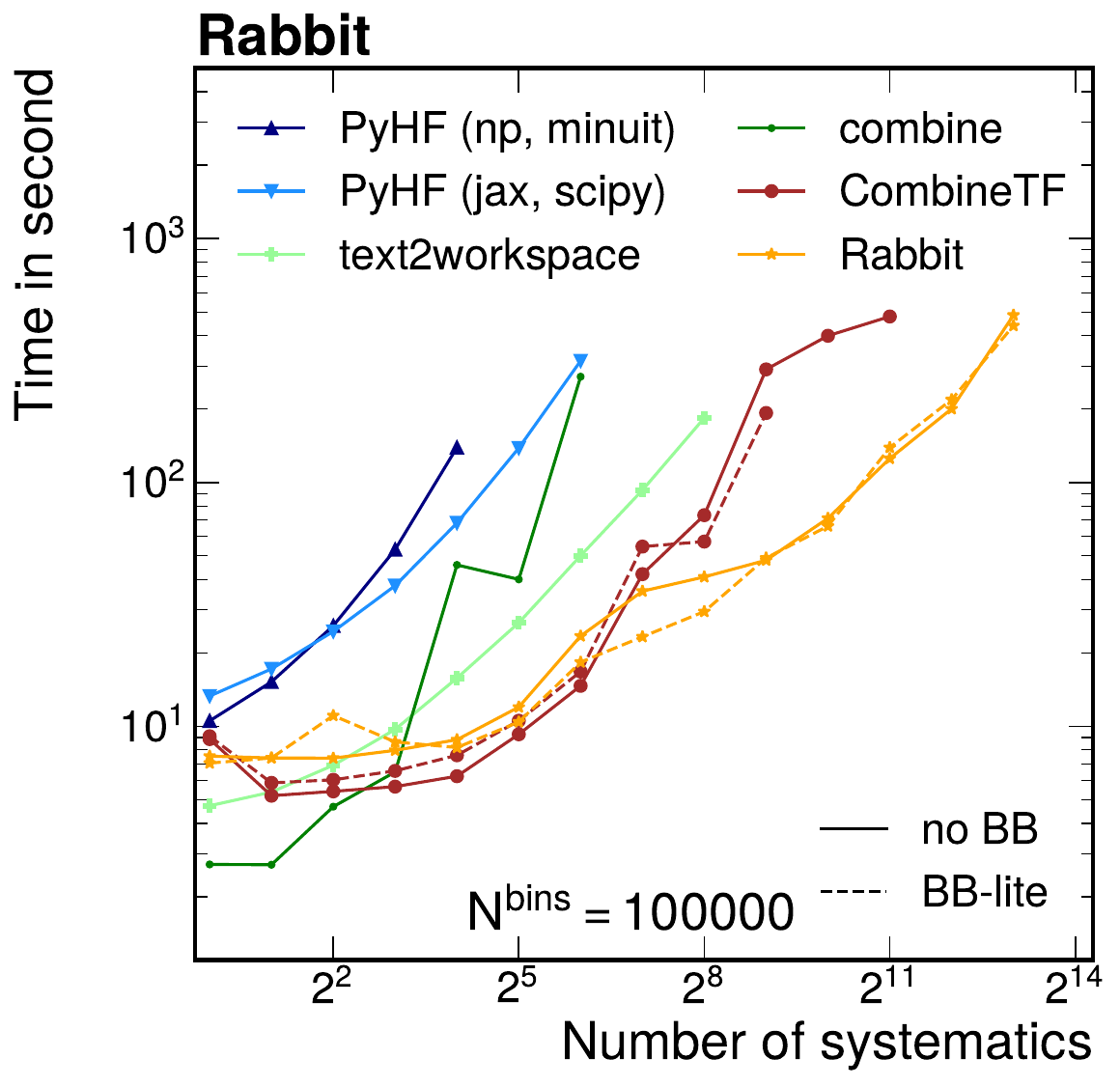}
\caption{Fit time, including uncertainty extraction, for models with 1\,000 (left) and 100\,000 (right) bins versus the number of systematic uncertainties for \rabbit, \combinetf, \combine, and \pyhf, with (dashed) and without (solid) bin-by-bin statistical uncertainties.}
\label{fig:comparison}
\end{figure}

Figure~\ref{fig:comparison} compares the time for minimization and uncertainty extraction with \pyhf~\cite{pyhf} (with \textsc{NumPy}+\textsc{Minuit} and with JAX+L-BFGS), \combine~\cite{combine} (including the datacard-to-workspace conversion), and \combinetf, the \tensorflow~1.x based predecessor of \rabbit developed for Ref.~\cite{whelicity}. For simple models all tools finish in under 10 seconds, with \combine fastest; as complexity increases, \pyhf and \combine exhibit steep growth in runtime, while \rabbit shows a much slower and delayed increase: at $\nbins = 1\,000$ and $\nsyst = 1\,024$ it completes in under 10 seconds where the standard tools exceed the 10-minute limit, and at $\nbins = 10\,000$ and $\nsyst = 4\,096$ it finishes in under one minute with no successful completion from the standard tools. Bin-by-bin statistical uncertainties strongly affect \pyhf, which gains $\nbins$ explicit parameters, while the other tools treat them analytically; the conversion step of \combine, however, becomes prohibitively expensive at high $\nbins$, and at $\nbins = 100\,000$ only \rabbit succeeds.

\section{Summary}
\label{sec:summary}

\rabbit combines differentiable programming with advanced statistical methods for efficient binned profile maximum likelihood estimation. Automatic differentiation, JIT compilation, strict vectorization, a Krylov-subspace trust-region minimizer, and analytic solutions where possible enable fits with unprecedented numbers of bins and nuisance parameters on CPUs and GPUs, outperforming established frameworks in regimes where they fail to converge within reasonable time. The framework has been used, among others, for the CMS W boson mass measurement~\cite{mw}, and such capabilities will be essential to fully exploit the data of the HL-LHC and beyond.

\begin{footnotesize}
\setlength{\bibsep}{1pt}

\end{footnotesize}

\end{document}